\documentclass[prl,twocolumn,aps,superscriptaddress,floatfix,tightenlines]{revtex4-2}
\usepackage{amsmath}
\usepackage{amsthm}
\usepackage{amssymb}
\usepackage{amsfonts}
\usepackage{epsfig}
\usepackage[dvips]{color}
\usepackage{bm}
\usepackage{times}
\usepackage[T1]{fontenc}
\usepackage[colorlinks,bookmarks=false,citecolor=blue,linkcolor=magenta,urlcolor=blue]{hyperref}

\begin{document}

\title{Metrological detection of multipartite entanglement through dynamical symmetries}
\author{Yu-Ran Zhang}
\affiliation{Theoretical Quantum Physics Laboratory, Cluster for Pioneering Research, RIKEN, Wakoshi, Saitama, 351-0198, Japan}
\affiliation{Quantum Computing Center, RIKEN, Wakoshi, Saitama, 351-0198, Japan}


\author{Franco Nori}
\email{fnori@riken.jp}
\affiliation{Theoretical Quantum Physics Laboratory, Cluster for Pioneering Research, RIKEN, Wakoshi, Saitama, 351-0198, Japan}
\affiliation{Quantum Computing Center, RIKEN, Wakoshi, Saitama, 351-0198, Japan}
\affiliation{Physics Department, University of Michigan, Ann Arbor, MI 48109-1040, USA}


\begin{abstract}
  Multipartite entanglement, characterized by the quantum Fisher information (QFI),
  plays a central role in quantum-enhanced metrology and
  understanding
  quantum many-body physics.
  With a dynamical generalization of the Mazur-Suzuki relations, we provide a rigorous lower bound
  on the QFI for the thermal Gibbs states in terms of
  dynamical symmetries, i.e., operators with periodic time dependence. We demonstrate
  that this bound
  can be saturated when considering a complete set of
  dynamical symmetries. Moreover, this lower bound
  with dynamical symmetries can be generalized to the QFI matrix and to the QFI for the thermal pure
  states, predicted by the eigenstate thermalization hypothesis.
  Our results reveal a new perspective to detect multipartite entanglement and other
  generalized variances in an
  equilibrium system, from its nonstationary dynamical properties, and is
   promising for studying emergent nonequilibrium many-body physics.
\end{abstract}

\maketitle

\emph{Introduction.---}%
Quantum Fisher information (QFI) is a central quantity in quantum metrology
that makes highly sensitive estimations of physical parameters by using
entanglement and quantum squeezing \cite{BRAUNSTEIN1994,Giovannetti2011,Ma2011,Pezze2018}.
By witnessing the quantum advantage in quantum metrology, the QFI detects metrologically
useful entanglement \cite{Toth2012,Guehne2009,Pezze2009}, reveals the hidden structure of multipartite entanglement
\cite{Ren2021,Xu2022} and identifies resourceful quantum states in general quantum resource theories \cite{Tan2021}.
However, to obtain the QFI is a strenuous task, even in a quantum many-body system with either exact solvability or
based on numerical simulations \cite{Jarzyna2013,Chabuda2020}.
Fortunately, the QFI can be determined from linear-response functions \cite{Shitara2016} and
 obtained through the dynamic susceptibility for thermal ensembles \cite{Hauke2016},
which are measurable in condensed-matter experiments \cite{Mathew2020}.
Corresponding to the finite-frequency dynamical response functions, the QFI
can be used to probe symmetry-breaking phenomena, such as quantum phase
transitions \cite{Hauke2016,Pezze2017,Zhang2018,Zhang2022a}.
In addition to the equilibrium systems, the non-ergodic physics
from different mechanisms have been attracting growing interests, including the many-body
localization (MBL) \cite{Basko2006,Znidaric2008,Kjaell2014,Schreiber2015,Smith2016,Xu2018,Lukin2019,Abanin2019},
Hilbert space fragmentation \cite{Sala2020,Bucva2022},
and quantum many-body scarred (QMBS) models
\cite{Bernien2017,Turner2018,Choi2019,Serbyn2021,Zhang2023,Desaules2022,Windt2022}.
As an experimentally extractable entanglement witness \cite{Strobel2014,Lu2020b,Yu2021},
the QFI is also used to study various out-of-equilibrium quantum many-body physics, e.g.,
 the MBL \cite{Smith2016,Guo2021}, the
 states in the eigenstate thermalization hypothesis (ETH) \cite{Brenes2020}, and
the QMBS states \cite{Desaules2022}.

The dynamical response functions depict the system's dynamical properties.
For instance,
ideal conductivity at finite temperatures is related to the Drude weight corresponding to the
\emph{zero-frequency} response function \cite{Zotos1997}.
This anomalous transport behavior can be explained using the Mazur-Suzuki
relations \cite{Mazur1969,Suzuki1971a} on the
 long-time average of the auto-correlation function of observables, in terms of equilibrium
 correlators involving conserved quantities \cite{Mazur1969}.
Moreover, the non-vanishing \emph{finite-frequency} dynamical response functions imply the
breaking of time-translation symmetry of thermal states by displaying the non-stationary dynamics.
These ergodicity-breaking dynamical phenomena have been related to the emergence of
extensive and local dynamical symmetries \cite{Buca2019,Medenjak2020a,Bucva2022}.
Similarly, a lower bound on the finite-frequency dynamical response functions has been
proposed in terms of dynamical symmetries \cite{Medenjak2020}. Thus, it is straightforward
to study the QFI through the dynamical symmetries.

In this work, 
we derive rigorous lower bounds on the QFI for thermal ensembles and thermal pure states
with dynamical symmetries. 
The lower bounds can be saturated given a complete set of
dynamical symmetries, which can be
generalized to the QFI matrices, Wigner-Yanase skew information, and the quantum variance.
Our results can be applied to detect multipartite entanglement
in several complex many-body systems with only limited knowledge of dynamical symmetries.
Since several nontrivial non-stationary dynamics are
related to the emergence of dynamical symmetries, our work
reveals a new viewpoint to investigate the entanglement structure of an
equilibrium system from its non-stationary dynamical properties by checking
whether the dynamical symmetries are strictly local or not.
Conversely, our results are promising for investigating out-of-equilibrium many-body
systems from their multipartite entanglement structures,
such as QMBS models, periodically driven integrable quantum systems, and time crystals.


\emph{Preliminaries.---}%
The QFI quantifies the distinguishability of a state $\rho$ from
a unitary transformed state
$\rho(\theta)=e^{-i\hat{\mathcal{O}}\theta}\rho e^{i\hat{\mathcal{O}}\theta}$, with a
Hermitian generator $\hat{\mathcal{O}}$ for an infinitesimal $\theta$ \cite{Giovannetti2011,Pezze2018}.
It constrains the achievable statistical precision
in quantum phase estimation by the quantum Cram\'{e}r-Rao bound (QCRB)
\cite{BRAUNSTEIN1994} as
$(\Delta\theta)^2\geq1/(\nu \mathcal{F}_Q)$, with $\nu$ being the repetition number of independent measurements. Given a decomposition of a mixed state $\rho=\sum_np_n|n\rangle\langle n|$ with $\langle m|n\rangle=\delta_{m,n}$,
the QFI with respect to the generator $\hat{\mathcal{O}}$ can be written as \cite{BRAUNSTEIN1994}
\begin{align}\label{qfisld}
\mathcal{F}_Q(\hat{\mathcal{O}})=\sum_{p_n+p_m>0}\frac{2(p_n-p_m)^2}{p_n+p_m}|\langle n|\hat{\mathcal{O}}|m\rangle|^2.
\end{align}
In addition, the QFI detects the metrologically useful multipartite entanglement,
when the QCRB can fall below the standard quantum limit, if $\rho$ is
entangled for $N$ particles. Concretely, for a local generator
$\hat{\mathcal{O}}=\sum_{\lambda}\hat{{o}}_\lambda$,
with $\hat{o}_\lambda$ having a spectrum of unit width,
the QFI for $\rho$ fulfilling $\mathcal{F}_Q/N>\kappa$ indicates
that $\rho$ is at least $(\kappa+1)$-partite entangled \cite{Hauke2016}.

Furthermore, the QFI for thermal mixed states at arbitrary temperatures is experimentally measurable with many-body correlations contained in the dynamical response functions \cite{Hauke2016,Shitara2016}.
Consider a thermal Gibbs state $\rho_\beta\equiv \sum_np_n{(\beta)}|E_n\rangle\langle E_n|$ of a system at an inverse temperature $\beta=1/T$,
 with $p_n{(\beta)}\equiv e^{-\beta E_n}/Z_\beta$ and $Z_\beta\equiv\sum_{m}e^{-\beta E_m}$.
Here, $|E_n\rangle$ denotes the eigenstate of the Hamiltonian $H$ for the eigenenergy $E_n$ with $H|E_n\rangle=E_n|E_n\rangle$.
The QFI for  $\rho_\beta$ with respect to $\hat{\mathcal{O}}$ can be expressed
in terms of Kubo response functions: \cite{Hauke2016}.
\begin{subequations}
\begin{align}
\mathcal{F}_Q(\hat{\mathcal{O}})&=\frac{1}{\pi}\int_{-\infty}^\infty\!\!\!\!d\omega\;\tanh({\beta\omega}/{2})\chi''_{\hat{\mathcal{O}}}(\omega)\label{QFI1}\\
&=\frac{1}{\pi}\int_{-\infty}^\infty\!\!\!\!d\omega\;\tanh^2({\beta\omega}/{2})\mathcal{S}_{\hat{\mathcal{O}}}(\omega).\label{QFI2}
\end{align}
\end{subequations}
Here,
$\chi''_{\hat{\mathcal{O}}}(\omega)\equiv \textrm{Im}\{{i}\int_{-\infty}^\infty dt e^{i\omega t}\langle[\hat{\mathcal{O}}(t),\hat{\mathcal{O}}]\rangle\}$ denotes the
imaginary part of the dynamic susceptibility, and
$\mathcal{S}_{\hat{\mathcal{O}}}(\omega)\equiv\int_{-\infty}^\infty dte^{i\omega t}(\langle\{\hat{\mathcal{O}}(t),\hat{\mathcal{O}}\}\rangle-2\langle\hat{\mathcal{O}}(t)\rangle\langle\hat{\mathcal{O}}\rangle)$
denotes the dynamic structure factor, $\hat{\mathcal{O}}(t)=e^{iHt}\hat{\mathcal{O}}e^{-iHt}$.
The average is taken on the Gibbs thermal state $\langle\circ\rangle\equiv\textrm{Tr}[\circ\rho_\beta]$.
Equations~(\ref{QFI1},\ref{QFI2}) are connected through the fluctuation-dissipation theorem:
$\chi''_{\hat{\mathcal{O}}}(\omega)=\tanh({\beta\omega}/{2})\mathcal{S}_{\hat{\mathcal{O}}}(\omega)$.

\begin{figure}[t]
	\centering
	\includegraphics[width=0.47\textwidth]{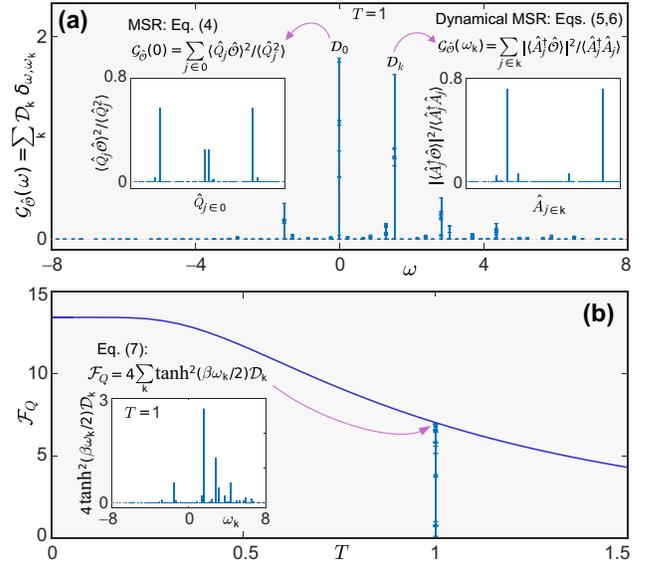}\\
	\caption{(a)  Dynamical response function $\mathcal{G}_{\hat{\mathcal{O}}}(\omega)$
  versus the frequency $\omega$, which exhibits the Mazur-Suzuki relation (MSR) at  zero frequency
  and the dynamical MSR for a nonzero frequency, in a spin-$\frac{1}{2}$ XX chain with a length $N=7$.
  Inset left: An illustration of the MSR as shown in Eq.~(\ref{smb}).
  Inset right: An illustration of the dynamical MSR as summarized in Eqs.~(\ref{dmb},\ref{dmb2}).
 	(b) Quantum Fisher information (QFI) versus temperature $T$.
  Inset: An illustration of the QFI in terms of dynamical symmetries, as summarized in Eq.~(\ref{tqfi}), at $T=1$.
}\label{fig:2}
\end{figure}

\emph{Rigorous lower bound of the QFI at arbitrary temperatures through dynamical symmetries.---}%
We first introduce a lower bound of the QFI through dynamical symmetries using a
rigorous lower bound of the dynamical response functions, which is proved in Ref.~\cite{Medenjak2020}.
The finite-frequency dynamical response function with $\omega\neq0$
is lower bounded as \cite{Medenjak2020}
\begin{align}
    \mathcal{G}_{\hat{\mathcal{O}}}(\omega)\equiv\lim_{\tau\rightarrow\infty}\frac{1}{2\tau}\int_{-\tau}^{\tau}\!\!\!\!dt\; e^{i\omega t}\langle\hat{\mathcal{O}}(t)\hat{\mathcal{O}}\rangle\geq\mathbb{A}^\dag\mathbb{V}^{-1}\mathbb{A},\label{df}
    \end{align}
%
where
$\mathbb{A}_j\equiv\langle \hat{A}_{j}^\dag\hat{\mathcal{O}}\rangle\delta_{\omega,\omega_{k\ni j}}$ and
$\mathbb{V}_{i,j}\equiv\langle\hat{A}^\dag_{i}\hat{A}_{j}\rangle$
are expressed in terms of a set of dynamical symmetries $\{\hat{A}_{j}\}$.
A dynamical symmetry is defined as an eigenoperator $\hat{A}_j$
of the Hamiltonian $H$, fulfilling
$[\hat{A}_j,H]=\omega_{\mathsf{k}\ni j}\hat{A}_j$, where $j\in \mathsf{k}$ denotes that the
dynamical symmetry $\hat{A}_j$ has an eigenfrequency $\omega_\mathsf{k}$.
The (zero-frequency) response function
$\mathcal{G}_{\hat{\mathcal{O}}}(0)$ reduces to the long-time-average
of the autocorrelation function, which is lower bounded by the Mazur-Suzuki relations \cite{Mazur1969,Suzuki1971a,Dhar2021}:
\begin{align}\label{smb}
\mathcal{G}_{\hat{\mathcal{O}}}(0)\geq\sum_{j\in0}{\langle \hat{Q}^{\bot}_{j}\hat{\mathcal{O}}\rangle^2}/\langle(\hat{Q}^{\bot}_j)^2\rangle\equiv\mathcal{D}_0(\hat{\mathcal{O}}),
\end{align}
with a set of orthogonal \emph{conserved quantities} $\{\hat{Q}^{\bot}_j\}$,
see the left inset in Fig.~\ref{fig:2}(a).
Here, we let $\omega_{\mathsf{k}=0}=0$, corresponding to the conserved quantities
with $[H,\hat{Q}^{\bot}_j]=0$.
In comparison, Eq.~(\ref{df}) shows that the \emph{dynamical} response function
has a rigorous lower bound in terms of \emph{dynamical symmetries}.

%

A set of $L$ linearly independent dynamical symmetries
$\{\hat{A}_{j}\}$ can be divided into $\mathsf{K}$ subsets corresponding to
 different nonzero frequencies, respectively.
It is obvious that $L=\sum_{\mathsf{k}=1}^{\mathsf{K}}l_\mathsf{k}$,
with each subset $\{\hat{A}_{j\in \mathsf{k}}\}$ having  $l_\mathsf{k}$
linearly independent dynamical symmetries with the same frequency $\omega_{\mathsf{k}}$.
Due to the time-translation
invariance of the thermal state, $\langle\hat{A}_i^\dag\hat{A}_j\rangle_\beta=0$ for $\omega_{\mathsf{k}\ni i}\neq\omega_{\mathsf{k}'\ni j}$.
Therefore, $\mathbb{V}$ is a block diagonal matrix:
$\mathbb{V}=\textrm{diag}(\mathbb{V}_1,\cdots,\mathbb{V}_\mathsf{K})$,
with nontrivial elements $\mathbb{V}_{i,j\in \mathsf{k}}=\langle\hat{A}_i^\dag\hat{A}_j\rangle$ in the $l_\mathsf{k}\times l_\mathsf{k}$
block $\mathbb{V}_\mathsf{k}$.
In addition,  $\mathbb{V}$ is Hermitian and can be diagonalized by a unitary
transformation $\hat{A}^{\bot}_{j\in k}=\sum_{i\in{k}}\mathbb{U}_{j,i}\hat{A}_i$ 
as
  $\mathbb{U}^\dag\mathbb{V}\mathbb{U}=\textrm{diag}(\langle (\hat{A}^{\bot}_1)^\dag\hat{A}^{\bot}_1\rangle,\cdots,\langle (\hat{A}^{\bot}_L)^\dag\hat{A}^{\bot}_L\rangle)$,
with $\mathbb{U}=\textrm{diag}(\mathbb{U}_1,\cdots,\mathbb{U}_\mathsf{K})$ and
$\mathbb{U}_\mathsf{k}^\dag\mathbb{U}_\mathsf{k}=\mathbb{I}_\mathsf{k}$.
Therefore, the lower bound (\ref{df}) for $\omega\neq0$
can be simplified  as
  \begin{align}\label{dmb}
    \mathcal{G}_{\hat{\mathcal{O}}}(\omega)\geq\pi\sum_\mathsf{k}\mathcal{D}_\mathsf{k}({\hat{\mathcal{O}}})\delta_{\omega,\omega_\mathsf{k}}, 
  \end{align}
which is a dynamical version of the Mazur bound with
\begin{align}\label{dmb2}
\mathcal{D}_\mathsf{k}(\hat{\mathcal{O}})\equiv\sum_{j\in \mathsf{k}}|\langle (\hat{A}^{\bot}_{j})^\dag\hat{\mathcal{O}}\rangle|^2/\langle(\hat{A}^{\bot}_j)^\dag\hat{A}^{\bot}_j\rangle.
\end{align}
In each term, $\mathcal{D}_\mathsf{k}$ corresponds to a subset of $l_k$
linearly independent dynamical symmetries
$\{\hat{A}_{j\in \mathsf{k}}\}$ with the same frequency $\omega_\mathsf{k}$, see the right inset in Fig.~\ref{fig:2}(a).

Combining Eqs.~(\ref{QFI2}) and (\ref{dmb}),
we have that the QFI for thermal ensembles is lower bounded as \cite{sm}
  \begin{align}\label{tqfi}
    \mathcal{F}_Q(\hat{\mathcal{O}})\geq\sum_{\mathsf{k}}{4}\tanh^2({\beta\omega_\mathsf{k}}/{2})\mathcal{D}_\mathsf{k}(\hat{\mathcal{O}}),
  \end{align}
where each term
corresponds to a subset of dynamical symmetries $\{\hat{A}_{j\in \mathsf{k}}\}$ with the same frequency $\omega_\mathsf{k}$. This lower bound is a direct application
of the dynamical Mazur bound (\ref{dmb}) that relates to the dynamical symmetries.
Moreover,  the QFI is completely irrelevant to the conserved quantities,
corresponding to the stationary dynamical behaviors of the system,
due to $\tanh(0)^2=0$.

\begin{figure}[t]
	\centering
	\includegraphics[width=0.47\textwidth]{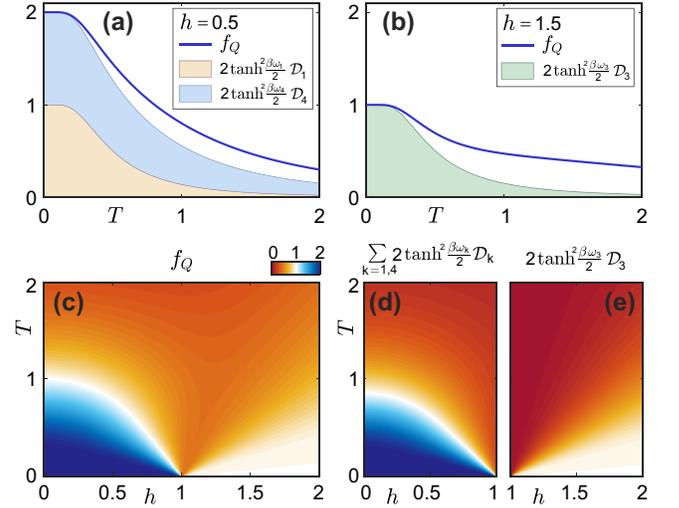}\\
	\caption{Metrologically useful entanglement detected through dynamical symmetries in a
	two-spin-$\frac{1}{2}$ XX model under an external field with a Hamiltonian
	$H_{\textrm{XX}}=(\hat{\sigma}_1^x\hat{\sigma}_{2}^x+\hat{\sigma}_1^y\hat{\sigma}_{2}^y)+h(\hat{\sigma}_{1}^z+\hat{\sigma}_{2}^z)$,
	given the generator $\hat{\mathcal{O}}=(\hat{\sigma}_1^x-\hat{\sigma}_{2}^x)/2$. (a) For $h=0.5$, the QFI density
	$f_Q=\mathcal{F}_Q/2$ is compared with the lower bound  with respect to two
	dynamical symmetries $\hat{A}_{1}=\hat{\sigma}_1^--\hat{\sigma}_1^z\hat{\sigma}_2^-$ and
	$\hat{A}_{4}=\hat{\sigma}_1^++\hat{\sigma}_1^z\hat{\sigma}_2^+$ versus the temperature $T$.
	(b) For $h=1.5$, the $f_Q$ is compared with the lower bound with respect to
	$\hat{A}_{3}=\hat{\sigma}_1^-+\hat{\sigma}_1^z\hat{\sigma}_2^-$ versus $T$.
	(c--e) The $f_Q$ (c) is compared with $\sum_{\mathsf{k}=1,4}2\tanh^2({\beta\omega_\mathsf{k}}/{2})\mathcal{D}_\mathsf{k}$ in the
	regime $h<1$ (d)  and  $2\tanh^2({\beta\omega_3}/{2})\mathcal{D}_3$   in the
	regime $h>1$ (e),  for different values of $T$.
}\label{fig:1}
\end{figure}

For example, we consider a two-qubit system with a XX interaction under a longitudinal field,
with a Hamiltonian
$H_{\textrm{XX}}=(\hat{\sigma}_1^x\hat{\sigma}_{2}^x+\hat{\sigma}_1^y\hat{\sigma}_{2}^y)+h(\hat{\sigma}_{1}^z+\hat{\sigma}_{2}^z)$, where $\hat{\sigma}^{x,y,z}$ are Pauli matrices.
We can find a complete set of dynamical symmetries:
$\hat{A}_1=\hat{\sigma}_1^--\hat{\sigma}_1^z\hat{\sigma}_2^-$,
$\hat{A}_2=\hat{A}_1^\dag$,
$\hat{A}_3=\hat{\sigma}_1^-+\hat{\sigma}_1^z\hat{\sigma}_2^-$, and
$\hat{A}_4=\hat{A}_3^\dag$, with frequencies: $\omega_{1,2}=\mp2(1+h)$, and
$\omega_{3,4}=\pm2(1-h)$. It is easy to verify that $\mathcal{F}_Q=\sum_{\mathsf{k}=1}^{4}4\tanh^2({\beta\omega_\mathsf{k}}/{2})\mathcal{D}_\mathsf{k}$.
In the regime $h<1$, when letting $t=1$ given the generator
$\hat{\mathcal{O}}=(\hat{\sigma}_1^x-\hat{\sigma}_{2}^x)/2$
[see Fig.~\ref{fig:1}(a,d)], the QFI density $f_Q\equiv\mathcal{F}_Q/2$ larger than $1$
 detects the presence of entanglement. A lower bound,
$\sum_{\mathsf{k}=1,4}2\tanh^2({\beta\omega_\mathsf{k}}/{2})\mathcal{D}_\mathsf{k}$, corresponding to two dynamical symmetries
$\hat{A}_{1,4}$, efficiently characterizes entanglement at low temperatures.
In the regime $h>1$ [see Fig.~\ref{fig:1}(b,e)], the lower bound, $2\tanh^2({\beta\omega_3}/{2})\mathcal{D}_3$,
with respect to $\hat{A}_{3}$, approximates to $f_Q$ at low temperatures.

\emph{Saturation of the lower bound on the QFI for thermal ensembles.---}%
Here we show that this lower bound can be saturated by considering
a complete set of dynamical symmetries.
We limit our study to the integrable model, whose Hamiltonian is exactly solvable by
the diagonalization $H=\sum_{n}E_n|E_n\rangle\langle E_n|$, with $\langle E_m|E_n\rangle=\delta_{m,n}$.
The non-integrable system with a finite size can be diagonalized numerically and investigated
in a similar way. Hereafter, we assume the non-degenerate case and leave the discussion
of the degeneracy in the Supplementary Materials (SM) \cite{sm}.

Without loss of generality, we consider a trivial complete set of orthogonal dynamical symmetries
$\{\hat{A}_{mn}\}$, whose elements are
$\hat{A}_{mn}=|E_m\rangle\langle E_n|$,
with $[H,\hat{A}_{mn}]=\omega_{mn}\hat{A}_{mn}$ and $\omega_{mn}=E_m-E_n$.
In this context, the inequality (\ref{dmb}) can be saturated,
which is the dynamical version of the Mazur-Suzuki relations.
Furthermore, with $\langle \hat{A}_{mn}^\dag\hat{\mathcal{O}}\rangle=p_n\langle E_m|\hat{\mathcal{O}}|E_n\rangle$
and $\langle\hat{A}_{mn}^\dag\hat{A}_{mn}\rangle=p_n$, we obtain from Eq.~(\ref{tqfi}) that
\begin{align}
  \mathcal{F}_Q
  &\geq
  2\sum_{mn}\tanh^2\frac{\beta\omega_{mn}}{2}|\langle E_m|\hat{\mathcal{O}}|E_n\rangle|^2 (p_n+p_m)=\mathcal{F}_Q,\nonumber
  \end{align}
where  $\tanh({\beta\omega_{mn}}/{2})={(p_n-p_m})/ ({p_n+p_m})$
and Eq.~(\ref{qfisld}) have been used for the equality.
Therefore, the inequality (\ref{tqfi}) can be saturated for a
complete set of dynamical symmetries, i.e.,  the
QFI for thermal ensembles can be divided to several terms, 
where in each term ${\mathcal{D}}_\mathsf{k}$ corresponds to a complete subset of dynamical symmetries with
 frequency $\omega_\mathsf{k}$, see Fig.~\ref{fig:2}(b).
%
%
The emergence of dynamical symmetries has been successfully applied to investigate
the dynamics of various non-stationary systems \cite{Buca2019,Medenjak2020a,Bucva2022}.
According to Eq.~(\ref{tqfi}), the generic multipartite entanglement structure of the thermal
Gibbs state, witnessed by the QFI, directly relates to
the non-stationary dynamical properties of the system, described by the dynamical symmetries.
Thus, our results provide a new perspective to investigate the multipartite entanglement for
thermal ensembles from the dynamical properties
of the system. Conversely, it would be useful for investigating
ergodicity-breaking phenomena in many-body systems, such as
 quantum many-body scars \cite{Desaules2022,Windt2022}, periodically driven integrable quantum systems \cite{Ljubotina2019}, and time crystals
 \cite{Medenjak2020a,Bucva2022}.

The lower bound (\ref{tqfi}) also provides a method to detect
the multipartite entanglement structure for several complex many-body systems,
e.g., the anisotropic Heisenberg  model \cite{Medenjak2020}, of which
only limited knowledge of extensive dynamical symmetries are known.
Moreover, since the nonzero QFI requires that
$\langle \hat{A}_{j}^\dag\hat{\mathcal{O}}\rangle_\beta\neq0$, it helps to search for the
effective generator to detect multipartite entanglement, according
to the forms of dynamical symmetries.
 In contrast, the strictly local dynamical symmetries, which
 would imply local Hilbert space fragmentation and out-of-time-ordered crystals \cite{Bucva2022},
 are ineffective for detecting the extensive multipartite entanglement.
For convenience, we assume the strictly local dynamical symmetries with forms
 $\hat{A}_{\textrm{loc}}=\hat{I}_{\lambda-1}\otimes\hat{a}_{{(\lambda,\lambda')}} \otimes\hat{I}_{N-\lambda'}$,
 where $\hat{a}_{(\lambda,\lambda')}$ only acts on $(r+1)$ particles on sites
 from $\lambda$ to $\lambda'$, with $r\equiv |\lambda'-\lambda|$. It is simple with
 the Cauchy-Schwarz inequality to prove that
${|\langle \hat{A}_{\textrm{loc}}^\dag\hat{\mathcal{O}}\rangle|^2}/{\langle\hat{A}_{\textrm{loc}}^\dag\hat{A}_{\textrm{loc}}\rangle}
\leq\langle(\sum_{\lambda\leq\nu\leq\lambda'}\hat{o})^2\rangle\leq{(r+1)^2}/{4}$,
 where $||o_\lambda||^2=1/4$, and without loss of generality,
  $\langle \hat{o}_\lambda\rangle=0$ is assumed.
  Therefore, the lower bound (\ref{tqfi})
  with  only strictly local dynamical symmetries
  cannot detect an extensive multipartite entanglement structure.



\emph{Extension to the Wigner-Yanase skew information and the quantum variance.---}%
Moreover, our results can be straightforwardly applied to study the Wigner-Yanase skew information
\cite{Wigner1963,Shitara2016}
and the quantum variance  \cite{Frerot2016,Frerot2019}, which also quantitatively relate to
the dynamical response functions. With the definition
$I_\alpha(\hat{\mathcal{O}})\equiv-\textrm{tr}\{[\hat{\mathcal{O}},{\rho}^{\alpha}_\beta][\hat{\mathcal{O}},{\rho}^{1-\alpha}_\beta]\}/2
$, the skew information
is lower bounded as
\begin{align}
I_{\frac{1}{2}}(\hat{\mathcal{O}})\geq\sum_{\mathsf{k}}\left[1-\frac{1}{\cosh(\beta\omega_\mathsf{k}/2)}\right]\mathcal{D}_\mathsf{k},
\end{align}
and the quantum variance, $(\Delta\hat{\mathcal{O}})_Q^2\equiv\int_0^1\!d\alpha I_\alpha(\hat{\mathcal{O}})$, can be limited as
\begin{align}
(\Delta\hat{\mathcal{O}})_Q^2\geq\sum_{\mathsf{k}}\left[1-\frac{\tanh(\beta\omega_\mathsf{k}/2)}{\beta\omega_\mathsf{k}/2}\right]\mathcal{D}_\mathsf{k}.
\end{align}
In fact, our results can be applied to estimate various measures and generalized covariances
of quantum fluctuations of the generator $\hat{\mathcal{O}}$,
which relates to the response functions through the fluctuation-dissipation
theorem \cite{Shitara2016}.

\emph{Generalization to the QFI matrix.---}%
The results can also be generalized to the QFI matrix \cite{Liu2020} for evaluating
the performance of quantum-enhanced multiple phase estimation \cite{Humphreys2013,Yue2014,Zhang2014}.
Consider $d$ parameters $\bm{\theta}\equiv(\theta_1,\cdots,\theta_d)$, imprinted on the thermal state by a unitary transformation
$\rho_\beta(\bm{\theta})=\exp(-i\hat{\bm{\mathcal{O}}}\bm{\theta})\rho_\beta\exp(-i\hat{\bm{\mathcal{O}}}\bm{\theta})$, with $\hat{\bm{\mathcal{O}}}=(\hat{\mathcal{O}}_{1},\cdots,\hat{\mathcal{O}}_d)$, where $\hat{\mathcal{O}}_a$ is the Hermitian generator for $\theta_a$
with $[\hat{\mathcal{O}}_a,\hat{\mathcal{O}}_b]=0$ for $a\neq b$.
The elements of the QFI matrix can be written as \cite{Gill2000,Shitara2016,Liu2020}
\begin{align}
[\mathbb{F}_Q]_{a,b}=\frac{1}{\pi}\int_{-\infty}^\infty\!\!\!\!d\omega\;\tanh^2({\beta\omega}/{2})\textrm{Re}[\mathcal{S}_{a,b}(\omega)],
\end{align}
where $\mathcal{S}_{a,b}(\omega)\equiv\int_{-\infty}^\infty dte^{i\omega t}[\langle\{\hat{\mathcal{O}}_a(t),\hat{\mathcal{O}}_b\}\rangle-2\langle\hat{\mathcal{O}}_a(t)\rangle\langle\hat{\mathcal{O}}_b\rangle]$
denotes the cross-correlation dynamic  structure factor.
Using the generalization of the lower bound (\ref{df}) to two different operators \cite{Medenjak2020}: (Here, $\phi$ is chosen
to make $e^{i\phi}\mathcal{G}_{a,b}(\omega)$ real.)
\begin{align}
e^{i\phi}\mathcal{G}_{a,b}(\omega)\geq e^{i\phi}\mathbb{A}_a^\dag\mathbb{V}^{-1}\mathbb{A}_b,
\end{align}
with $[\mathbb{A}_{a}]_j\equiv\langle \hat{A}_{j}^\dag\hat{\mathcal{O}}_{a}\rangle\delta_{\omega,\omega_{k\ni j}}$, the QFI matrix can be divided for a complete set of dynamical symmetries for all eigen-frequencies as \cite{sm}
\begin{align}\label{qfim}
\mathbb{F}_Q(\bm{\mathcal{O}})
    =\sum_{\mathsf{k}}{4}\tanh^2({\beta\omega_\mathsf{k}}/{2})\textrm{Re}[{\mathbb{D}}_\mathsf{k}(\hat{\bm{\mathcal{O}}})],
\end{align}
where ${[\mathbb{D}}_{\mathsf{k}}]_{a,b}\equiv\sum_{j\in \mathsf{k}}\langle (\hat{A}^{\bot}_{j})^\dag\hat{\mathcal{O}}_a\rangle\langle \hat{\mathcal{O}}_b\hat{A}^{\bot}_{j}\rangle/\langle(\hat{A}^{\bot}_j)^\dag\hat{A}^{\bot}_j\rangle$
in each term corresponds to a complete subset of dynamical symmetries with
the same frequency $\omega_\mathsf{k}$.

 \emph{Lower bound for the QFI in the ETH.---}%
%
Since the dynamical Mazur-Suzuki relations in Eqs.~(\ref{dmb},\ref{dmb2})
hold for different types of ensembles in statistical mechanics,
our results can be extended to the study of the QFI
for thermal pure states, fulfilling the ETH.
The ETH ansatz for the 
generator
$\hat{\mathcal{O}}$ in the basis
of the eigenstates of the Hamiltonian $H$
can be formulated as \cite{Srednicki1999,DAlessio2016}
$\mathcal{O}_{m,n}=\mathcal{O}({E})\delta_{m,n}+e^{-S({E})/2}g_{\hat{\mathcal{O}}}(\omega,{E})\mathbb{R}_{m,n}$, 
where ${E}\equiv (E_m+E_n)/2$, $\omega\equiv E_m-E_n$, $S(E)$
is the thermodynamic entropy at energy $E$, 
$\mathcal{O}(E)$ is the expectation value of the microcanonical
ensemble at ${E}$, $g_{\hat{\mathcal{O}}}(\omega,{E})$ is
a smooth function, and
$\mathbb{R}_{m,n}$ is a random variable with zero mean and unit variance.
In this context, it is shown in Ref.~\cite{Brenes2020}  that
the QFI of a thermodynamic ensemble  with a pure eigenstate at $\beta$
for the generator $\hat{\mathcal{O}}$ satisfying the ETH is written as
$\mathcal{F}^{\textrm{ETH}}_Q=\int_{-\infty}^\infty d\omega\mathcal{S}_{\hat{\mathcal{O}}}(\omega)/\pi$,
where
$\mathcal{S}_{\hat{\mathcal{O}}}(\omega)\simeq4\pi\cosh(\beta\omega/2)|g_{\hat{\mathcal{O}}}(\omega,{E})|^2$ \cite{Brenes2020}. 
The inverse temperature, defined as $\beta=\partial S(E)/\partial E$,
corresponds to the canonical temperature at energy ${E}=\textrm{Tr}(e^{-\beta H}H)/Z_\beta$.
The average 
is taken on
a pure state $|E\rangle$ at the effective $\beta$ with energy $E$ for the generator
$\hat{\mathcal{O}}$ satisfying the ETH.
%
%

Similarly, using  both the static and dynamical versions of the Mazur-Suzuki relations,
we can obtain that \cite{sm}
\begin{align}
     \mathcal{F}_Q^{\textrm{ETH}}(\hat{\mathcal{O}})\gtrsim\sum_{\mathsf{k}\neq0}{4}\mathcal{D}_\mathsf{k}(\hat{\mathcal{O}}).\label{ethqfi}
   \end{align}
Here, $\mathsf{k}\neq0$ implies that the QFI in the ETH only corresponds to
{dynamical symmetries} rather than conserved quantities.
Furthermore, the dynamic structure factor
$\mathcal{S}_{\hat{\mathcal{O}}}(\omega)$, evaluated from the ETH,
approximates its canonical counterpart \cite{Brenes2020}.
The difference between the QFI calculated on thermal ensembles and the one in the ETH could be evaluated as
\begin{align}
\mathcal{F}_Q^{\textrm{ETH}}(\hat{\mathcal{O}})-\mathcal{F}_Q(\hat{\mathcal{O}})\gtrsim\sum_{\mathsf{k}\neq0}{4}\mathcal{D}_\mathsf{k}(\hat{\mathcal{O}})/{\cosh^{2}({\beta\omega_\mathsf{k}}/{2})},
\end{align}
when $\mathcal{D}_\mathsf{k}(\hat{\mathcal{O}})$ is considered
on thermal ensembles.

\emph{Conclusions and discussions.---}%
We demonstrate a rigorous lower bound on the QFI for a thermal equilibrium system
through its dynamical symmetries, which characterize the nonstationary dynamical
properties of the system. 
Our results can be extended to the generalized (co)variances of the generators
that relate to the linear-response functions, including the QFI matrices,
Wigner-Yanase skew information, and the quantum variance.
The inequality is tight when we consider a complete set of dynamical symmetries,
so the QFI can be divided into different terms, each of which corresponds
to a complete subset of dynamical symmetries with the same frequency.
As a result, we can investigate the structure of multipartite entanglement of
a quantum many-body system  and find the appropriate generator for the entanglement witness
from the forms of the dynamical symmetries.
 Since the emergence of dynamical symmetries has been used to explain various
nontrivial non-stationary dynamics \cite{Buca2019,Medenjak2020a,Bucva2022},
our work provides a new angle to study the generic multipartite entanglement
structure that relates to  the non-stationary dynamical properties of the system.
Furthermore, our results can be extended to investigate various emergent
ergodicity-breaking dynamical phenomena, such as
 QMBS models \cite{Bernien2017,Turner2018,Choi2019,Serbyn2021,Zhang2023,Desaules2022,Windt2022}, Floquet driven quantum systems \cite{Grifoni1998,Ljubotina2019,Medenjak2020},
 and time crystals
 \cite{Else2016,Khemani2016,Yao2017,Else2017,Zhang2017b,Choi2017,Machado2020,Randall2021,Kyprianidis2021,Mi2022,Frey2023}.

\begin{acknowledgements}
We thank Tomotaka Kuwahara and Henning Schomerus for valuable discussions.
This work is supported in part by:
Nippon Telegraph and Telephone Corporation (NTT) Research,
the Japan Science and Technology Agency (JST) [via
the Quantum Leap Flagship Program (Q-LEAP), and
the Moonshot R\&D Grant No.~JPMJMS2061],
the Army Research Office (ARO) (Grant No.~W911NF-18-1-0358),
the Asian Office of Aerospace Research and Development (AOARD) (via Grant No.~FA2386-20-1-4069), and
the Foundational Questions Institute Fund (FQXi) via Grant No.~FQXi-IAF19-06.
\end{acknowledgements}

\bibliography{Manuscriptl.bib}

\newpage

\end{document}


\title{Supplementary Material for \\
Metrological detection of multipartite entanglement through dynamical symmetries}

\author{Yu-Ran Zhang}
\affiliation{Theoretical Quantum Physics Laboratory, Cluster for Pioneering Research, RIKEN, Wakoshi, Saitama, 351-0198, Japan}
\affiliation{Quantum Computing Center, RIKEN, Wakoshi, Saitama, 351-0198, Japan}


\author{Franco Nori}
\email{fnori@riken.jp}
\affiliation{Theoretical Quantum Physics Laboratory, Cluster for Pioneering Research, RIKEN, Wakoshi, Saitama, 351-0198, Japan}
\affiliation{Quantum Computing Center, RIKEN, Wakoshi, Saitama, 351-0198, Japan}
\affiliation{Physics Department, University of Michigan, Ann Arbor, MI 48109-1040, USA}
\maketitle
\beginsupplement


\section{Proof of the dynamical Mazur-Suzuki relations for thermal ensembles}
For $\omega\neq0$, the dynamical response function in Eq.~(3) in the main text
can be calculated as
\begin{align}
    \mathcal{G}_{\hat{\mathcal{O}}}(\omega)&=\lim_{\tau\rightarrow\infty}\frac{1}{2\tau}\int_{-\tau}^{\tau}\!\!\!\!dt\; e^{i\omega t}\langle\hat{\mathcal{O}}(t)\hat{\mathcal{O}}\rangle\nonumber\\
    &=\sum_{mn}p_n{(\beta )}\lim_{\tau\rightarrow\infty}\frac{1}{2\tau}\int_{-\tau}^{\tau}\!\!\!\!dt\; e^{i(\omega-\omega_{mn}) t}|\langle E_m|\hat{\mathcal{O}}|E_n\rangle|^2\nonumber\\
    &=\sum_{mn}p_n{(\beta )}|\langle E_m|\hat{\mathcal{O}}|E_n\rangle|^2\delta_{\omega,\omega_{mn}}
    \end{align}
where  $\omega_{mn}=E_m-E_n$, and we have used the fact that
\begin{align}
\lim_{\tau\rightarrow\infty}\frac{1}{2\tau}\int_{-\tau}^{\tau}\!\!\!\!dt\; e^{i(\omega-\omega_{mn}) t}=\delta_{\omega,\omega_{mn}}
\end{align}
with $\delta_{\omega,\omega_{mn}}$ being the Kronecker delta from the infinite-long-time average.

Given a trivial complete set of orthogonal dynamical symmetries
$\{\hat{A}_{mn}=|E_m\rangle\langle E_n|\}$, with $[H,\hat{A}_{mn}]=\omega_{mn}\hat{A}_{mn}$ and $\langle\hat{A}_{mn}^\dag\hat{A}_{m'n'}\rangle=p_n{(\beta )}\delta_{m,m'}\delta_{n,n'}$, we can obtain that
\begin{align}
&\langle \hat{A}_{mn}^\dag\hat{\mathcal{O}}\rangle=p_n{(\beta )}\langle E_m|\hat{\mathcal{O}}|E_n\rangle \delta_{\omega,\omega_{mn}},\\
&\langle\hat{A}_{mn}^\dag\hat{A}_{mn}\rangle =p_n{(\beta )},
\end{align}
and further
\begin{align}
\mathcal{D}_{mn}({\hat{\mathcal{O}}})&=\frac{|\langle \hat{A}_{mn}^\dag\hat{\mathcal{O}}\rangle|^2}{\langle \hat{A}_{mn}^\dag\hat{A}_{mn}\rangle^2}
=\sum_{mn}p_n{(\beta )}|\langle E_m|\hat{\mathcal{O}}|E_n\rangle|^2,
\end{align}
which closes the proof of the dynamical Mazur-Suzuki relations:
\begin{align}
\mathcal{G}_{\hat{\mathcal{O}}}(\omega)=\sum_{mn}\mathcal{D}_{mn}({\hat{\mathcal{O}}})\delta_{\omega,\omega_{mn}}.\label{MZR}
\end{align}

The dynamical response function for two different Hermitian generators $\hat{\mathcal{O}}_a$
and $\hat{\mathcal{O}}_b$ can be written as
\begin{align}
    \mathcal{G}_{a,b}(\omega)&=\lim_{\tau\rightarrow\infty}\frac{1}{2\tau}\int_{-\tau}^{\tau}\!\!\!\!dt\; e^{i\omega t}\langle\hat{\mathcal{O}}_a(t)\hat{\mathcal{O}}_b\rangle\\
    &=\sum_{mn}p_n{(\beta )}\langle E_m|\hat{\mathcal{O}}_a|E_n\rangle\langle E_m|\hat{\mathcal{O}}_b|E_n\rangle\delta_{\omega,\omega_{mn}}.
\end{align}
Then, with
\begin{align}
{[\mathbb{D}}_{mn}]_{a,b}&=\frac{\langle \hat{A}_{mn}^\dag\hat{\mathcal{O}}_a\rangle\langle \hat{\mathcal{O}}_b\hat{A}_{mn}\rangle}{\langle\hat{A}_{mn}^\dag\hat{A}_{mn}\rangle}\\
&=p_n{(\beta )}\langle E_m|\hat{\mathcal{O}}_a|E_n\rangle\langle E_n|\hat{\mathcal{O}}_b|E_m\rangle,
\end{align}
we have
\begin{align}\label{dsuzukiequality}
\mathcal{G}_{a,b}(\omega)=\sum_{mn}{[\mathbb{D}}_{mn}]_{a,b}\delta_{\omega,\omega_{mn}}.
\end{align}

\section{Proof of Equations~(10,14)}
For thermal states, the dynamic structure factor can be expressed using Eq.~(\ref{MZR}) as
\begin{align}
\mathcal{S}_{\hat{\mathcal{O}}}(\omega)&=\lim_{\tau\rightarrow\infty}\int_{-\tau}^\tau \!\!\!\!dt\;e^{i\omega t}(\langle\{\hat{\mathcal{O}}(t),\hat{\mathcal{O}}\}\rangle-2\langle\hat{\mathcal{O}}\rangle^2)\nonumber\\
&=\lim_{\tau\rightarrow\infty}2\tau[\mathcal{G}_{\hat{\mathcal{O}}}(\omega)+\mathcal{G}_{\hat{\mathcal{O}}}(-\omega)]-4\pi\langle\hat{\mathcal{O}}\rangle^2\delta({\omega})\nonumber\\
&\geq\sum_{\mathsf{k}}2\pi[\delta(\omega-\omega_\mathsf{k})+\delta(\omega+\omega_\mathsf{k})]\mathcal{D}_\mathsf{k}({\hat{\mathcal{O}}})-4\pi\langle\hat{\mathcal{O}}\rangle^2\delta({\omega}),\nonumber
\end{align}
where we have used the relationship between the Kronecker delta and the Dirac delta function as
\begin{align}
\lim_{\tau\rightarrow\infty}2\tau\delta_{\omega,\omega_{mn}}=2\pi\delta(\omega-\omega_{mn}).
\end{align}
Using Eq.~(2b) in the main text, we can express the quantum Fisher inforamtion
(QFI) as
\begin{align}
\mathcal{F}_Q(\hat{\mathcal{O}})
&=\frac{1}{\pi}\int_{-\infty}^\infty\!\!\!\!d\omega\;\tanh^2({\beta\omega}/{2})\mathcal{S}_{\hat{\mathcal{O}}}(\omega)\nonumber\\
&\geq
\sum_{\mathsf{k}\neq0}4\tanh^2({\beta\omega_\mathsf{k}}/{2})\mathcal{D}_\mathsf{k}({\hat{\mathcal{O}}}),
\end{align}
where the term corresponding to the conserved quantities
with $\mathcal{D}_{k=0}({\hat{\mathcal{O}}})$ and $\omega_{0}=0$
does not contribute to the QFI due to $\tanh(0)=0$.
For the last inequality, we have used  and the fact that
$\tanh^2(x)$ is an even function, and
\begin{align}
\tanh({\beta\omega_{mn}}/{2})=\frac{p_n{(\beta)}-p_m{(\beta)}}{{p_n{(\beta)}+p_m{(\beta)}}}.
\end{align}

For the QFI in the eigenstate thermalization hypothesis (ETH), for convenience,
we consider the average on the thermal ensembles, as
a proper approximation as discussed in Ref~\cite{Brenes2020}, and have
\begin{align}
\mathcal{F}^{\textrm{ETH}}_Q(\hat{\mathcal{O}})
&\simeq\frac{1}{\pi}\int_{-\infty}^\infty\!\!\!\!d\omega\;\mathcal{S}(\omega,\hat{\mathcal{O}})\nonumber\\
&\geq\sum_{\mathsf{k}\neq0}4\mathcal{D}_\mathsf{k}({\hat{\mathcal{O}}}) + 4[\mathcal{D}_0({\hat{\mathcal{O}}})-\langle\hat{\mathcal{O}}\rangle^2]\\
&=\sum_{\mathsf{k}\neq0}4\mathcal{D}_\mathsf{k}({\hat{\mathcal{O}}}).
\end{align}
For the last equality, we have used the Mazur-Suzuki relations
in Eq.~(4) in the main text,
when choosing all the energy projection operators $\{\hat{Q}_n=|E_n\rangle\langle E_n|\}$
as a trivial complete set of conserved quantities.

\section{Quantum Fisher information matrix for thermal ensembles through dynamical symmetries}
The cross-correlation on the thermal ensembles can be calculated as
\begin{align}
\mathcal{S}_{a,b}(\omega)&=\lim_{\tau\rightarrow\infty}\int_{-\tau}^\tau\!\!\!\! dt\;e^{i\omega t}(\langle\{\hat{\mathcal{O}}_a(t),\hat{\mathcal{O}}_b\}\rangle-2\langle\hat{\mathcal{O}}_a\rangle\langle\hat{\mathcal{O}}_b\rangle)\nonumber\\
&=\lim_{\tau\rightarrow\infty}2\tau[\mathcal{G}_{a,b}(\omega)+\mathcal{G}_{a,b}(-\omega)]-4\pi\langle\hat{\mathcal{O}}_a\rangle\langle\hat{\mathcal{O}}_b\rangle\delta{(\omega)}\nonumber\\
&=\sum_{mn}2\pi[\delta(\omega-\omega_{mn})+\delta(\omega+\omega_{mn})]{[\mathbb{D}}_{mn}]_{a,b}-4\pi\langle\hat{\mathcal{O}}\rangle^2\delta({\omega}),\nonumber
\end{align}
where for convenience, only the dynamical Suzuki equality (\ref{dsuzukiequality}) is used. Because $\mathcal{G}_{a,b}(\omega)$
would be a complex function [see Ref.~\cite{Medenjak2020} and Eq.~(13) in the main text],
 the elements of the QFI matrix can be written as
\begin{align}
[\mathbb{F}_Q]_{a,b}&=\frac{1}{\pi}\int_{-\infty}^\infty\!\!\!\!d\omega\;\tanh^2({\beta\omega}/{2})\textrm{Re}[\mathcal{S}_{a,b}(\omega)]\\
&=4\sum_{k}\tanh^2({\beta\omega}/{2})\textrm{Re}\{[\mathbb{D}_{mn}]_{a,b}\}
\end{align}
where we have used the fact that
\begin{align}
\textrm{Re}\{[\mathbb{D}_{mn}]_{a,b}\}&=\sum_{mn}\textrm{Re}[\langle \hat{A}_{mn}^\dag\hat{\mathcal{O}}_a\rangle\langle \hat{\mathcal{O}}_b\hat{A}_{mn}\rangle]/\langle\hat{A}_{mn}^\dag\hat{A}_{mn}\rangle\nonumber\\
&=\sum_{mn}\frac{p_n{(\beta )}+p_m{(\beta )}}{2}\textrm{Re}[\langle E_m|\hat{\mathcal{O}}_a|E_n\rangle\langle E_n|\hat{\mathcal{O}}_b|E_m\rangle],
\end{align}
and
\begin{align}
[\mathbb{F}_Q]_{a,b}&=2\sum_{mn}\frac{[p_n{(\beta)}-p_m{(\beta)}]^2}{p_n{(\beta)}+p_m{(\beta)}}\textrm{Re}[\langle E_m|\hat{\mathcal{O}}_a|E_n\rangle\langle E_n|\hat{\mathcal{O}}_b|E_m\rangle].
\end{align}

\section{Degenerate energy spectrum}
So far we have assumed a nondegenerate energy spectrum for the Hamiltonian, when proving the equality of
the dynamical Mazur-Suzuki relations as summarized in Eq.~(9) in the main text,
which is not necessary and greatly simplifies the notation.
Here, we reformulate the main results of the main text when considering a degenerate energy spectrum,
of which the Hamiltonian can be written as
\begin{align}
  H=\sum_{n}E_n\sum_{\mu=1}^{d_n}|E_n^{\mu}\rangle\langle E_n^\mu|
\end{align}
with  $d_n$ being the degeneracy of the energy level $E_n$ and
$\langle E_m^\nu|E_n^\mu\rangle=\delta_{m,n}\delta_{\mu,\nu}$.
We can write the thermal state as
\begin{align}
\rho_\beta&=\frac{1}{Z_\beta}\sum_{n}{e^{-\beta E_n}}\sum_{\mu=1}^{d_n}|E_n^\mu\rangle\langle E_n^\mu|\nonumber\\
&=\sum_{n}\sum_{\mu=1}^{d_n}p_n^\mu(\beta)|E_n^\mu\rangle\langle E_n^\mu|
\end{align}
with $Z_\beta\equiv \sum_{n}d_n{e^{-\beta E_n}}$ and $p_n^\mu{(\beta )}=e^{-\beta E_n}/(\sum_{n}d_n{e^{-\beta E_n}})$.

Similarly, we can also consider a trivial complete set of orthogonal dynamical symmetries
$\{\hat{A}^{\nu\mu}_{mn}\}$, of which the elements are
$\hat{A}^{\nu\mu}_{mn}=|E_m^{\nu}\rangle\langle E_n^{\mu}|$,
with $[H,\hat{A}^{\nu\mu}_{mn}]=\omega_{mn}\hat{A}^{\nu\mu}_{mn}$ and $\omega_{mn}=E_m-E_n$.
Then, we have
\begin{align}
\langle (\hat{A}^{\nu\mu}_{mn})^\dag\hat{\mathcal{O}}\rangle=p_n^\mu{(\beta )}\langle E_m^\nu|\hat{\mathcal{O}}|E_n^\mu\rangle
\end{align}
and
\begin{align}
  \langle(\hat{A}_{mn}^{\nu\mu})^\dag\hat{A}_{mn}^{\nu\mu}\rangle=p_n^\mu.
\end{align}
Then, we can obtain that
\begin{align}
{4}\sum_{mn,\mu\nu}\frac{|\langle (\hat{A}^{\nu\mu}_{mn})^\dag\hat{\mathcal{O}}\rangle|^2}{\langle(\hat{A}_{mn}^{\nu\mu})^\dag\hat{A}_{mn}^{\nu\mu}\rangle}\tanh^2\frac{\beta\omega_{mn}}{2}&={2}\sum_{mn,\mu\nu}\frac{[p_n^\mu{(\beta )}-p_m^\nu{(\beta )}]^2}{p_n^\mu{(\beta )}+p_m^\nu{(\beta )}}\langle E_m^\nu|\hat{\mathcal{O}}|E_n^\mu\rangle\\
  &
  =\mathcal{F}_Q(\hat{\mathcal{O}}),
\end{align}
which is the QFI for the thermal state with a degenerate energy spectrum.
Thus, our main results as shown in Eq.~(9) in the main text still hold for the thermal state of
a Hamiltonian having a degenerate energy spectrum, which can also be simply extended
to the Wigner-Yanase skew information, the quantum variance, and the QFI matrix.

\bibliography{Manuscriptl.bib}